\documentclass[review,12pt]{elsarticle}

\usepackage{amssymb}
\usepackage{amsmath}
\usepackage{xcolor}
\usepackage{lineno}
\usepackage{array}
\usepackage{makecell}
\usepackage{caption}
\usepackage{subcaption}
\usepackage{multirow}
\usepackage{lineno}
\usepackage{comment}
\usepackage{tabularx}
\usepackage[ragged]{sidecap}
\usepackage{siunitx,upgreek}
\usepackage{lineno}
\biboptions{sort&compress}

\usepackage{amsfonts}
\usepackage{graphicx}
\usepackage{setspace}
\usepackage{tocloft}
\usepackage{lipsum}
\usepackage{multirow}
\usepackage{xcolor}
\usepackage{comment}
\usepackage{soul}
\usepackage{url}

\usepackage{lineno}

\journal{Nuclear Instruments and Methods in Physics Research Section A}

\begin{document}

\begin{frontmatter}

\title{Beam Test Performance of AstroPix sensor with 120 GeV protons}

\author[a]{Bobae Kim}
\ead{bobae.kim@anl.gov}

\author[b]{Regina Caputo}
\author[a]{Manoj Jadhav}
\author[a]{Sylvester Joosten}
\author[b]{Carolyn Kierans}
\author[a]{Henry Klest}
\author[b,c]{Adrien Laviron}
\author[d]{Richard Leys}
\author[a]{Jessica Metcalfe}
\author[a]{Jared Richards}
\author[d]{Nicolas Striebig}
\author[e]{Amanda L. Steinhebel}
\author[b,c]{Daniel Violette}
\author[a]{Maria Żurek}
\cortext[cor1]{Corresponding author}

\affiliation[a]{organization={Argonne National Laboratory},addressline={9700 S. Cass Avenue}, city={Lemont},postcode={IL 60439}, state={Illinois}, country={U.S.A.}}
\affiliation[b]{organization={NASA Goddard Space Flight Center},addressline={8800 Greenbelt Rd},city={Greenbelt},postcode={MD 20771}, state={Maryland}, country={U.S.A.}}
\affiliation[c]{NASA Postdoctoral Program Fellow (ORAU)}
\affiliation[d]{organization={ASIC and Detector Laboratory, Karlsruhe Institute of Technology},addressline={Hermann-von-Helmholtz-Platz 1}, city={Karlsruhe}, postcode={D-76344}, city={Baden-Wurttemberg}, country={Germany}}
\affiliation[e]{organization={Oak Ridge National Laboratory},addressline={1 Bethel Valley Road Oak Ridge},city={Oak Ridge},postcode={TN 37830}, state={Tennessee}, country={U.S.A.}}
 
\begin{abstract}
AstroPix is a High-Voltage CMOS Monolithic Active Pixel Sensor (HV-CMOS MAPS) developed for precision gamma-ray imaging and spectroscopy in the medium-energy regime, as well as for precise shower imaging and tracking in the Barrel Imaging Calorimeter (BIC) of the Electron Proton/Ion Collider (ePIC) detector at the future Electron-Ion Collider (EIC).
We present beam test results of the AstroPix\_v3 sensor using a 120 GeV proton beam at the Fermilab Test Beam Facility (FTBF), performed as part of the broader experimental campaign for the BIC prototype calorimeter.
The sensor’s 500~\unit{\micro\meter} pixel pitch enabled precise measurement of the beam profile, providing important information for the calorimeter performance studies.
Using the measured 120 GeV proton data, we measure the energy deposit of minimum ionizing particles (MIP) and use them to extract the corresponding effective depletion depth at a single bias voltage of $-150~\mathrm{V}$.
\end{abstract}



\end{frontmatter}


\section{Introduction}
\label{sec:intro}

AstroPix has been developed for the proposed All-sky Medium-Energy Gamma-ray Observatory eXplorer (AMEGO-X) mission~\cite{amx_paper}, a next-generation MeV gamma-ray telescope building on the heritage of the Fermi Large Area Telescope (Fermi-LAT)~\cite{Atwood2009FermiLAT}.
AMEGO-X is designed to explore key questions in high-energy astrophysics, including the origins of cosmic rays and neutrinos, the physics of relativistic jets from neutron star mergers, the role of supermassive black holes, and the acceleration sites of galactic particles, while providing continuous full-sky monitoring of energetic transients.
AMEGO-X is composed of a tracker, based on AstroPix, and a cesium iodide calorimeter. 
The AstroPix-based tracker is undergoing performance evaluation and technology validation through near-space and sub-orbital flight demonstrations, including the Compton and pair (ComPair-2) gamma-ray telescope~\cite{compair2} and the AstroPix Sounding Rocket Technology dEmonstration Payload (A-STEP)~\cite{Violette2024ASTEP}.

In addition, AstroPix has been selected as the imaging layer for the Barrel Imaging Calorimeter (BIC) of the Electron Proton/Ion Collider (ePIC) detector at the future Electron-Ion Collider (EIC).
The EIC will provide high-luminosity polarized electron–proton and electron–ion collisions with unprecedented precision.
The ePIC experiment, the first experiment at the EIC, is dedicated to exploring fundamental questions in nuclear physics including the origin of the nucleon mass and the nucleon spin, as well as dense gluonic systems~\cite{WhitePaper, NASReport, YellowReport}.

The BIC is the electromagnetic calorimeter in the barrel region of the ePIC detector and identifies scattered electrons, a key requirement for accurate reconstruction of event kinematics.
The BIC consists of lead/scintillating-fiber (Pb/SciFi) sampling calorimeter layers interleaved with AstroPix-based imaging layers embedded in the front half of the calorimeter~\cite{ATHENA:2022hxb}.
This structure provides precise three-dimensional imaging of particle showers, enabling essential performance capabilities such as $e^- / \pi^\pm$ and $\gamma/\pi^0$ separation, which are crucial for Deep Inelastic Scattering (DIS) measurements and for achieving the EIC’s scientific goals~\cite{YellowReport}.

For the AstroPix imaging layers, one of the key requirements for the BIC is to achieve sufficient dynamic range to detect minimum ionizing particle (MIP) signals in all layers.
The dynamic range and energy resolution of the third iteration of the AstroPix chips, AstroPix\_v3, have been studied using radioactive gamma-ray sources in a previous study~\cite{SUDA2024169762}.

In this work, we extend these studies to a beam environment by testing AstroPix\_v3 with a 120 GeV proton beam at the Fermilab Test Beam Facility (FTBF)~\cite{ftbfBeamOverview}. 
Since 120 GeV protons behave as minimum ionizing particles, they provide an ideal beam to study the MIP response of AstroPix.
From 120 GeV proton data, we measure the energy deposited in the sensor and derive the corresponding effective depletion depth, which is a key parameter for accurate detector simulations and system-level performance studies.
Section 2 describes the specifications of AstroPix\_v3 (Section 2.1), the beam-test preparation and calibration (Section 2.2), and the experimental setup used in the FTBF test (Section 2.3).
Section 3 presents the beam test results, and Section 4 summarizes the conclusions and future prospects.

\begin{figure*}[h!]
    \centering
    \includegraphics[width=0.95\textwidth]{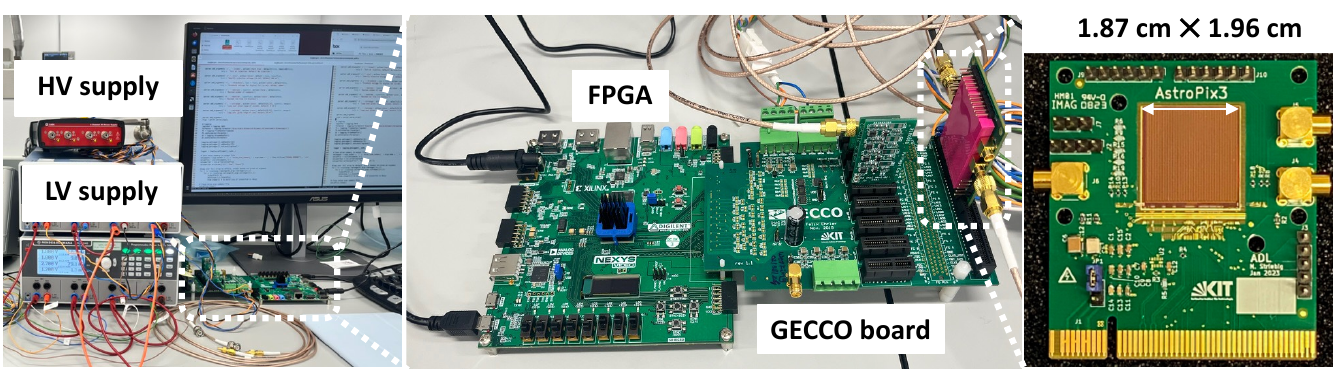}
    \caption{Bench test setup with AstroPix\_v3 single chip.}
    \label{fig:bench}
\end{figure*}
\section{Experimental Setup}
\subsection{AstroPix\_v3}
AstroPix is a novel High-Voltage CMOS Monolithic Active Pixel Sensor (HV-CMOS MAPS) that benefits from a deep depletion region enabled by a high bias voltage, allowing fast drift-based charge collection and improved radiation hardness compared to conventional MAPS~\cite{PERIC2007876}.
It features in-pixel amplification and in-pixel signal processing, as well as on-chip digitization, which minimizes signal routing and power consumption.
Furthermore, the detector chips can be daisy-chained to enable efficient scalability to large-area and multilayer detectors.

Since 2019, the development and testing of AstroPix has progressed through several design iterations, from version 1 to version 4~\cite{SUDA2024169762, Brewer:2021mbe, Steinhebel:2022ips, LSteinhebel:2023NO, steinhebel2025astropix, suda2026astropix4, striebig2024astropix4}. 
The on-going AstroPix Research and Development (R\&D) aims to meet the requirements of both AMEGO-X and the BIC.
The time resolution and power consumption requirements differ between the two systems, with AMEGO-X targeting approximately 1~\unit{\micro s} and below 1.5~\unit{mW/cm^2}, while the BIC requires about 20~ns and around 2~\unit{mW/cm^2}.
In addition, common goals for both applications include achieving an energy resolution better than 10\% at 122~keV (FWHM), a target dynamic range covering 25--700~keV, and operating in the fully depleted region to maximize charge collection.
The nominal depletion-depth specification for AstroPix for both AMEGO-X and the BIC applications is 500~\unit{\micro m}.
AstroPix\_v5 is designed to achieve full depletion depth through the use of high resistivity (> 5~\unit{k}$\Omega \cdot$\unit{\cm}), in combination with backside processing.

AstroPix\_v3, the first full-scale chip (1.87~cm~$\times$~1.96~cm), has been extensively evaluated for use in both the BIC imaging layers and NASA-led gamma-ray telescopes. 
AstroPix\_v3 comprises a 35 $\times$ 35 pixel matrix with a pixel pitch of 500~\unit{\micro\meter}.
Each pixel contains a 300 $\times$ 300~\unit{\micro\meter}$^2$ high-voltage deep n-well (DNW), which forms a depletion region at the DNW/p-substrate junction under reverse bias.
CMOS readout circuitry including a charge amplifier, CR-RC shaper, and comparator, is integrated within each DNW.
AstroPix\_v3 was fabricated using TSI Semiconductors’ 180~nm process, with a p-type substrate resistivity of 200--400~$\Omega \cdot$\unit{\cm} and thickness of 725~\unit{\micro\meter}.
Further details can be found in~\cite{steinhebel2025astropix}.

AstroPix\_v3 employs pixel-level self-triggering with streaming readout, allowing each pixel to autonomously record hits above threshold. The chip operates with a 2.5~MHz clock for time-of-arrival (ToA) measurement using an 8-bit counter and up to a 200~MHz clock for time-over-threshold (ToT) measurement using a 12-bit counter.

The sensor design incorporates row and column hit buffers, where hit information from a single pixel is recorded separately in the row and column streams due to the buffer architecture. Each hit in a row (resp. column) generates a data packet containing the chip ID, the row (resp. column) number, the ToA, and the ToT. 
The complete hit information for each pixel is reconstructed in post-processing by matching the corresponding row and column hit records. This matching is performed during the pre-analysis stage using the recorded row and column data, based on the following criteria:
(1) the absolute ToA difference is less than 2 clock cycles, meaning that the row and column ToA values are expected to be identical ($\Delta$ToA = 0), while occasional one-cycle differences ($\Delta$ToA = 1) are tolerated to account for timing jitter and readout latency;
(2) the relative ToT difference, defined as $(|ToT_{column}-ToT_{row}|)/ToT_{column}$ is required to be below 10\%.

\subsection{Beam test preparation and calibration}

As shown in Figure~\ref{fig:bench}, the AstroPix test setup consists of a customized GEneric Configuration and COntrol (GECCO~\cite{schimassek2021development}) data acquisition system, a Nexys Video board equipped with a Xilinx Artix-7 FPGA, and a carrier board used to integrate the chips into the GECCO system.
Bond pads on a single AstroPix\_v3 chip, located along the lower edge of the chip’s digital periphery, are directly wire-bonded to a rigid carrier PCB for electrical connection and communication via an SPI interface.
A bias voltage of $-150~\mathrm{V}$ is supplied using a DT5533EN desktop high-voltage power supply module from CAEN Technologies.
Analog (VDDA) and digital (VDDD) supply voltages of $1.8~\mathrm{V}$, along with a $2.7~\mathrm{V}$ GECCO power voltage, are provided by an HMP404 power supply (Rohde\&Schwarz) to power the AstroPix readout circuitry and the FPGA logic, respectively.

The bias voltage of $-150~\mathrm{V}$ was selected based on I--V measurements and on bench tests performed to optimize the VDAC settings, including threshold and dynamic range.
At $-150~\mathrm{V}$, the leakage current lies within the plateau region, indicating operation close to full depletion.
Higher bias voltages were found to increase noise fluctuations, which are particularly undesirable in a high-rate test-beam environment.
Therefore, $-150~\mathrm{V}$ was chosen as the nominal operating point to minimize noise while maintaining stable detector performance.

In preparation for the beam test, noise measurement and radiation source tests were conducted on the bench setup to prepare the experimental configuration.
AstroPix\_v3 employs a global threshold applied uniformly across all pixels. The first three columns were excluded because they use a different comparator design, which results in higher noise due to greater amplification.

\begin{figure}[h!]
    \centering
    \includegraphics[width=0.8\linewidth]{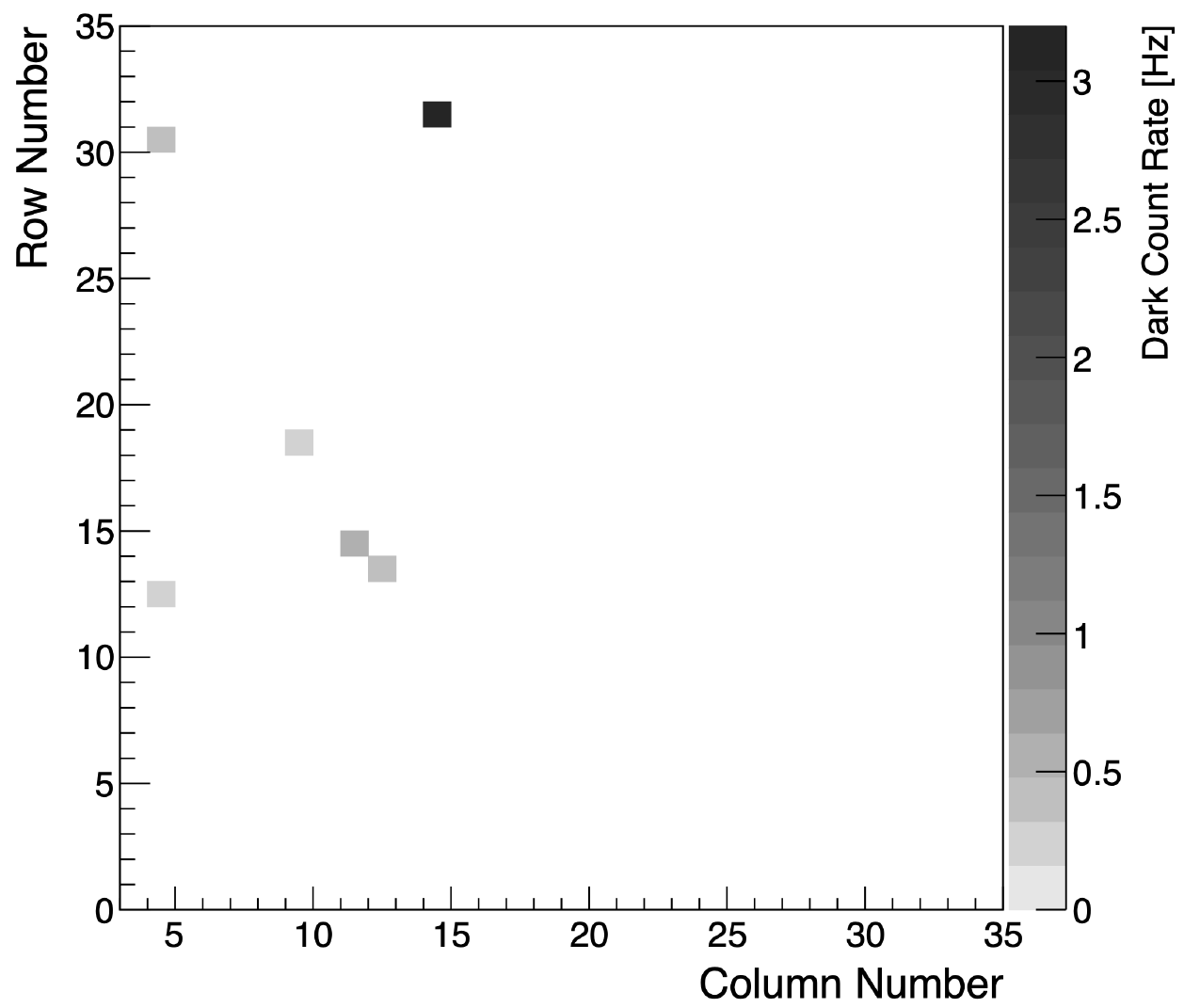}
    \caption{Noise map obtained at a 200 mV threshold, presenting the dark count rate across the pixel array.}
    \label{fig:noisemap}
\end{figure}

To determine an optimal threshold, the dark count rates of all pixels were measured as a function of threshold. The goal was to keep the dark count rate below 2~Hz while maintaining high active pixel yield. Based on the threshold scan, a threshold of 200~mV was selected, corresponding to approximately 20~keV based on the calibration curve. At this threshold, only one of the 1,120 pixels exceeded a dark count rate of 2~Hz, while the remaining pixels satisfied the target noise requirement. This threshold also satisfies the AstroPix dynamic range floor of 25~keV. Figure~\ref{fig:noisemap} presents the noise map at this threshold, illustrating the dark count rate across the pixel array.
This pixel located at (14, 31), which exceeds 2~Hz, is masked throughout the measurements.

\begin{figure}[h!]
    \centering
    \includegraphics[width=.9\linewidth]{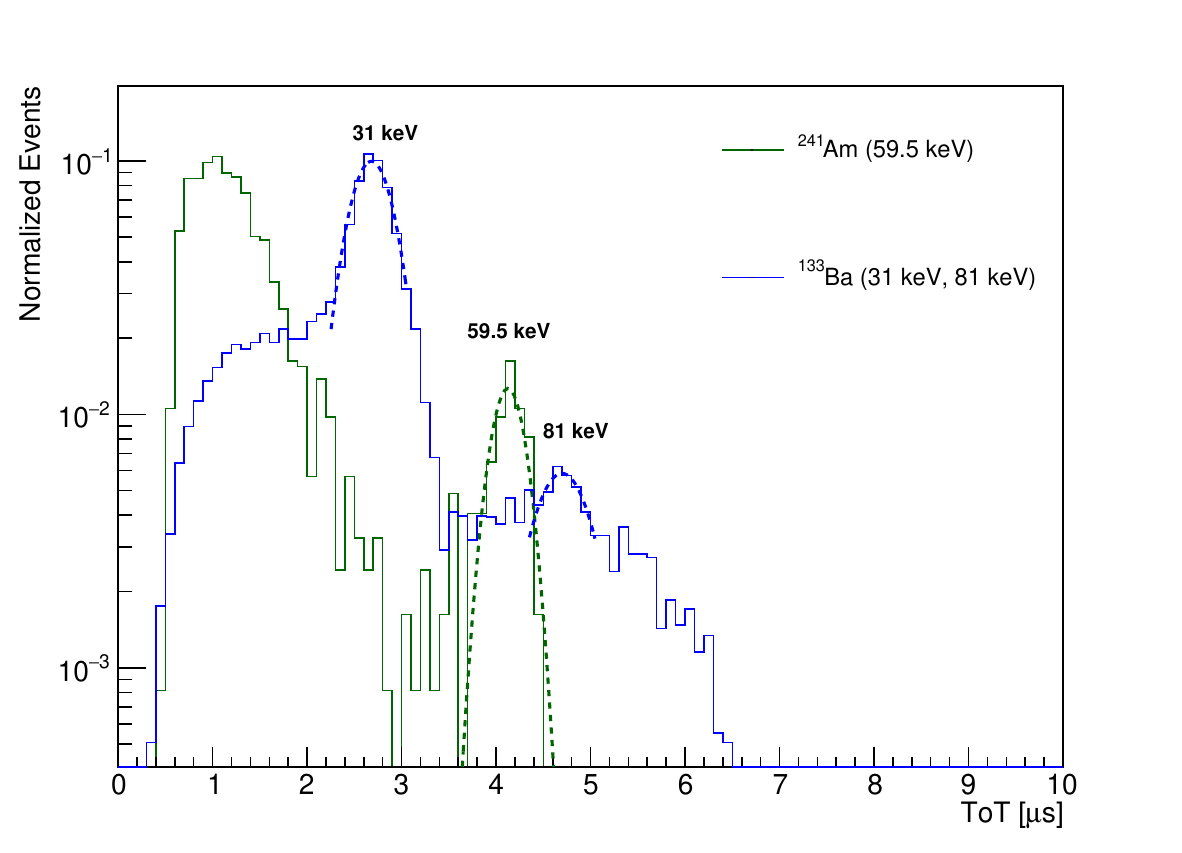}  
    \caption{One example of the single-pixel ToT spectra from the $^{241}$Am and $^{133}$Ba measurements is shown. The photopeaks are identified and fitted with Gaussian functions. The green and blue lines correspond to the $^{241}$Am and $^{133}$Ba spectra, respectively, and the dotted lines indicate the fitted Gaussian functions.} 
    \label{fig:totspecc16r3}
\end{figure}

The source test was performed using $^{241}$Am and $^{133}$Ba sources to obtain per-pixel energy calibration curves.
The ToT distribution per pixel from gamma rays of 59.5~keV from $^{241}$Am and 31~keV and 81~keV from $^{133}$Ba was measured  providing three calibration points.
These calibration points are sufficient to convert the measured ToT value into energy, as the deposited MIP energy is expected to lie between 31~keV and 59.5~keV based on the $^{90}$Sr energy measurements.

\begin{figure}[h!]
    \centering
    \includegraphics[width=0.99\linewidth]{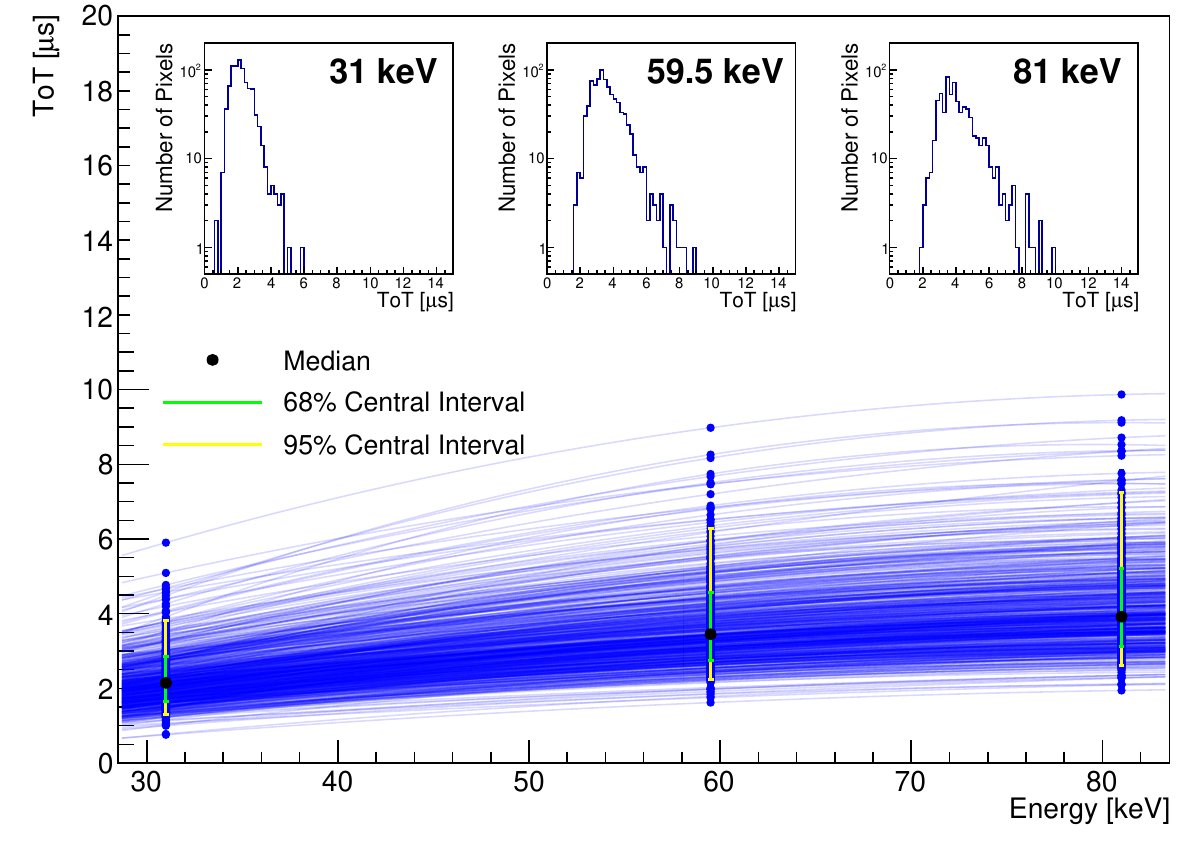}
    \caption{Calibration curves for all selected pixels of a single AstroPix\_v3 sensor. Blue lines show fits with the second-degree polynomial function to the mean ToT values extracted from Gaussian fits to the ToT distributions at three gamma energies (31, 59.5, and 81~keV). 
    The upper panels display the distributions of the mean ToT values across pixels for each energy, illustrating pixel-to-pixel variations. The green and yellow bands indicate the 68\% and 95\% central intervals around the median, respectively.}
    \label{fig:calibcurve}
\end{figure}

Figure~\ref{fig:totspecc16r3} shows example ToT spectra for one pixel of $^{241}$Am and $^{133}$Ba measurements, where the 31~keV, 59.5~keV, and 81~keV photopeaks are clearly visible and fitted with Gaussian functions.
Pixels with low statistics or without a visible photopeak were excluded from the calibration. Clear photopeaks of 31~keV, 59.5~keV, and 81~keV were observed in 1,101 pixels (98.3\%), 859 pixels (76.7\%), and 816 pixels (72.9\%), respectively, out of the 1,120 pixels under test.
Among these, 683 pixels (61.0\%) had clear photopeaks at all three gamma energies and were calibrated using a second-order polynomial function, as shown in Figure~\ref{fig:calibcurve}. 
The top panels of Figure~\ref{fig:calibcurve} show the distributions of the mean values obtained from Gaussian fits to the ToT spectra across all selected pixels for each energy point, illustrating pixel-to-pixel variation.
Central intervals, presented as 68\% in green and 95\% in yellow, are overlaid on all calibration curves to indicate the spread of pixel responses.

The fraction of pixels with identifiable photopeaks depends on several effects. At lower energies, pixel-to-pixel gain variations and the absence of pixel-by-pixel threshold tuning may affect the peak visibility. However, the 31 keV peak was still observed in 98.3\% of the pixels.
While the response at higher energies is primarily limited by the reduced effective dynamic range due to the nonlinear ToT response, only 72.9\% of the pixels show an identifiable 81 keV peak.
In addition, broadened energy resolution, overlapping spectral structures, signal spread, and saturation effects during signal propagation can further reduce the number of pixels for which the peaks can be reliably identified. Therefore, the calibration was performed using the subset of well-behaved pixels with clearly identifiable peaks at all three energies to evaluate optimal device performance.

Compared to the previous study~\cite{SUDA2024169762}, in which 89.6\% of the tested pixels were calibrated at $-350~\mathrm{V}$, the present study was performed at the nominal operating bias of $-150~\mathrm{V}$.
$-350~\mathrm{V}$ is close to breakdown and is mainly useful for extreme-bias studies, rather than representing the intended operating point for the applications considered here.
In addition, chip-to-chip variations in gain, noise, threshold dispersion, and dynamic-range behavior may also contribute to the different calibrated-pixel fractions.

The energy calibration function used in the previous study~\cite{SUDA2024169762}, which targeted calibration over a wide energy range, 
$ \mathrm{ToT}=ax+b[1-e^{-x/c}]+d $,
combining linear and exponential terms, was used to describe the measured non-linear sensor response from 22.2~keV to 122.1~keV. Instead, the present work focuses on the MIP region, rather than on calibration over the full dynamic range. The observed source peaks cover the energy range relevant for MIP signals and therefore provide the calibration points needed for this measurement. Given this narrower calibration range, simpler functional forms were considered, while the empirical function used in the previous wide-range study was retained as a reference for comparison.

To assess suitable alternative functions, an injection study was performed.
AstroPix employs an in-pixel test-pulse capacitor for charge injection. A digital pulse generates a voltage step defined by the VDAC setting, resulting in an injected charge $Q_{\mathrm{inj}} = C_{\mathrm{inj}} \Delta V_{\mathrm{inj}}$, which is proportional to the applied voltage step $V_{\mathrm{inj}}$.
A direct voltage injection spanning 100--800~mV is provided, which is used to evaluate and compare two different fitting functions relating injection voltage to ToT.

\begin{figure}[h!]
    \centering
    \includegraphics[width=0.8\linewidth]{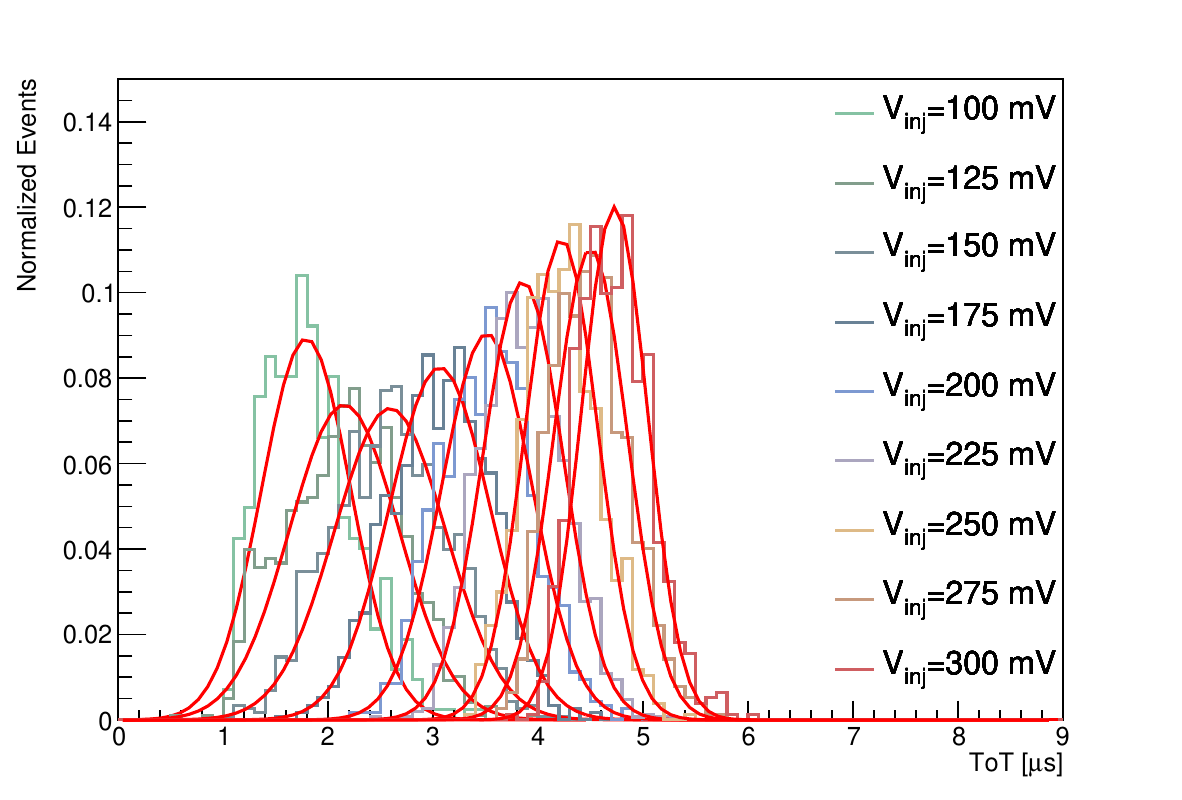}
    \includegraphics[width=0.8\linewidth]{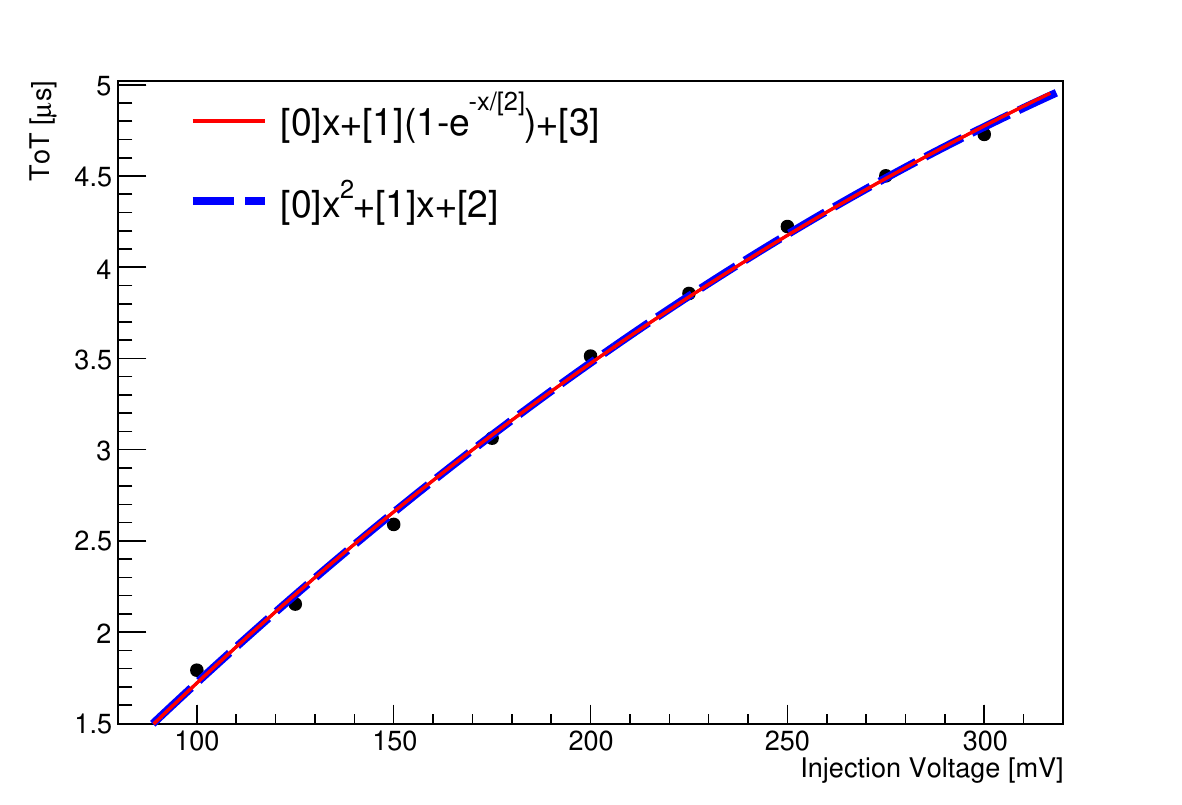}
    \caption{(Top) Example ToT spectra for injection voltages from 100~mV to 300~mV. All ToT distributions are fitted with a Gaussian function. (Bottom) Example fitting results for ToT value as a function of injection voltages using two fit functions: an empirical function (red line) and a second-degree polynomial function (blue dotted line).}
    \label{fig:injfit}
\end{figure}

The top plot of Figure~\ref{fig:injfit} shows example ToT spectra from a representative pixel for injection voltages between 100~mV and 300~mV in 25~mV steps, with all ToT distributions fit using a Gaussian function.
This voltage range was chosen because it spans the ToT values corresponding to the 31~keV, 59.5~keV, and 81~keV photopeaks observed in the same pixel.
The bottom plot of Figure~\ref{fig:injfit} presents example fits of the ToT value as a function of the injection voltage using two functional forms: the empirical function used in previous studies and the second-degree polynomial. 
The red solid line and blue dotted line represent the two fits, respectively. 
Both functions provide consistent results within the ToT range relevant for the 120~GeV proton measurement and the gamma-ray photopeaks.
Therefore, a second-degree polynomial function is used as the calibration function in this work.

The charge-injection calibration was not used as the primary calibration method in this work. 
For some pixels, MIP-region signals correspond to injection voltages below the minimum accessible injection range, preventing a reliable injection-based calibration. Therefore, the charge-injection study was used only as a cross-check and to guide the choice of calibration function, while the energy calibration was based on the measured source peaks.

\subsection{Beam Test at FTBF}
\label{sec:ftbf}
The beam test was conducted at the Fermilab Test Beam Facility (FTBF) in June 2024 using a 120 GeV proton beam.
This measurement was part of the same experimental campaign described in Ref.~\cite{Klest_2025}.
The AstroPix sensor was installed at the MTest beam line in the 6.2A enclosure~\cite{ftbfBeamOverview}, together with the Pb/SciFi prototype calorimeter for BIC. 

\begin{figure*}[h!]
    \centering
    \includegraphics[width=0.85\textwidth]{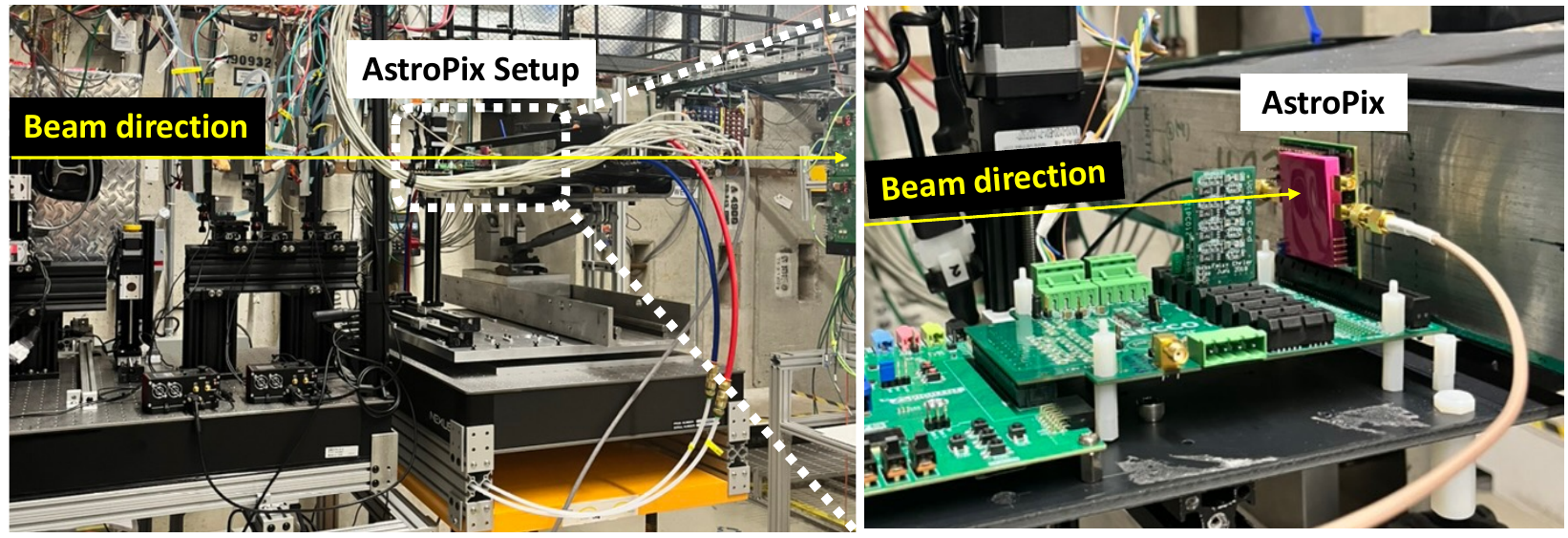}
        \caption{FBTF experimental setup with AstroPix\_v3 single chip.}
    \label{fig:ftbf}
\end{figure*}

As shown in Figure~\ref{fig:ftbf}, the AstroPix sensor was aligned with the geometric center of the beamline using the FTBF laser positioning system.
The sensor was oriented with its surface perpendicular to the beam axis and positioned approximately 2~cm upstream of the calorimeter face.
Since the multiwire proportional chambers typically used for tracking at FTBF were unavailable during this beam period, the AstroPix sensor was employed not only for its primary MIP response studies but also to provide the beam profile needed for the calorimeter study.

The MTest beamline is capable of delivering protons, muons, and mixed beams of electrons and pions, depending on the configuration of targets, absorbers, and collimators.
For this beam test, a 120 GeV primary proton beam, directly extracted from the main injector, was used for AstroPix testing.
The beam consists of 100\% protons and can be operated at the highest intensity of up to 500,000 particles per spill, with a spot size confined within 6 mm~\cite{ftbfBeamMTest}; however, this value is noted to depend strongly on the beam tune.

During the AstroPix test run, the proton beam was delivered in 4.2~s-long spills every 1 minute with about 55,000 particles per spill, resulting in a delivered particle rate of around 13 kHz within each spill. This beam intensity was chosen to provide a rate already above the expected maximum operating rate of approximately 1~kHz/chip for the BIC application, and also above the rates expected for space-based applications of AstroPix\_v3.

\section{Results}

\subsection{Beam Profile}

\begin{figure*}[h!]
\centering
\includegraphics[width=.4\linewidth]{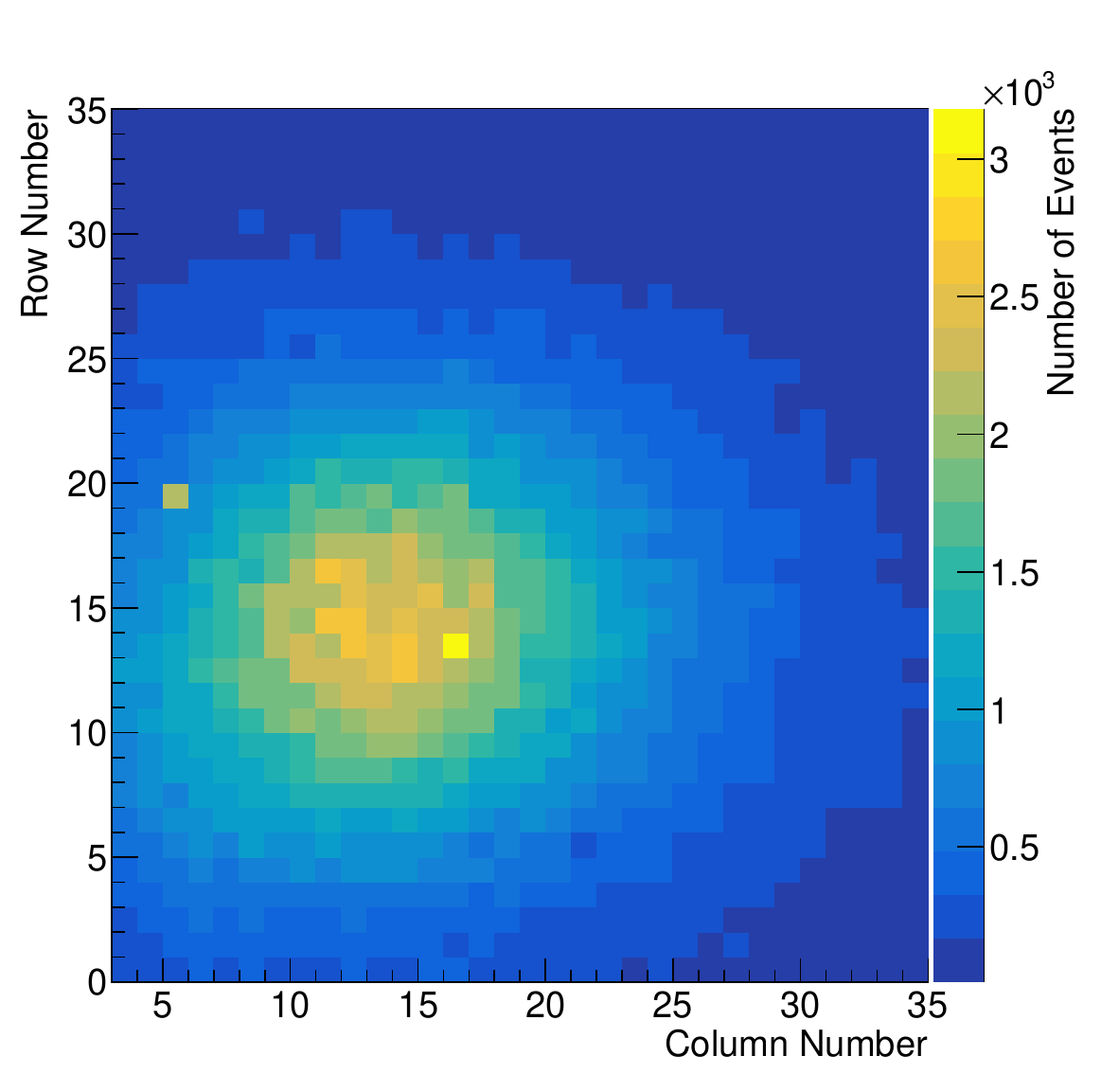}
\includegraphics[width=.4\linewidth]{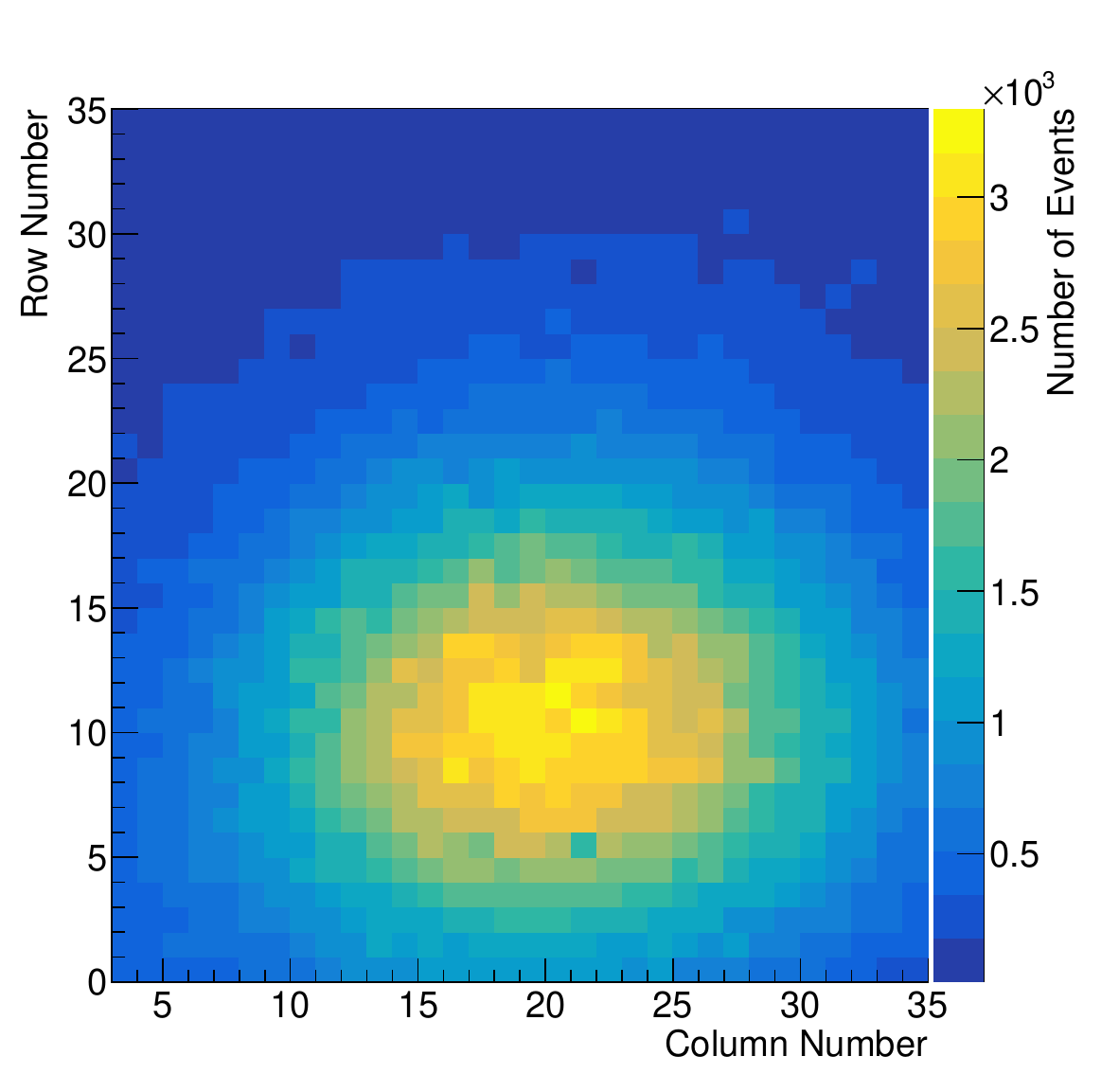}
\caption{Two-dimensional hit maps of 120~GeV protons for each run. Each map displays the number of matched hits in column–row coordinates of the AstroPix sensor.}
\label{fig:hitmaps}
\end{figure*}

\begin{figure*}[h!]
\centering
\includegraphics[width=1\linewidth]{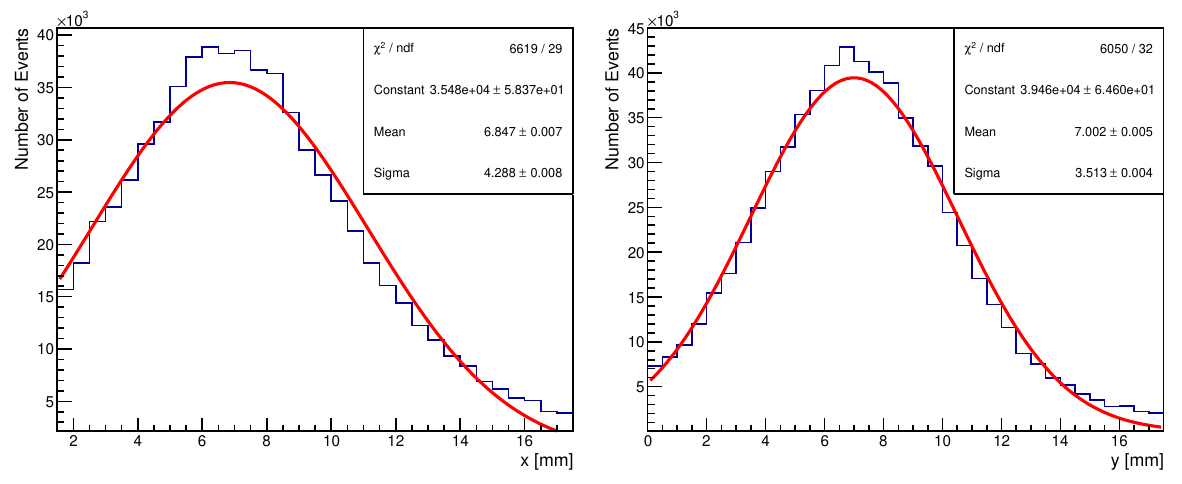}
\caption{Example beam profile of the first run in the horizontal (left) and vertical (right) directions, obtained by projecting the two-dimensional hit map onto the respective axes. 
Each histogram is fitted with a Gaussian function and the fit parameters are reported in the legend.}
\label{fig:beamprofile}
\end{figure*}

Figure~\ref{fig:hitmaps} presents the two-dimensional hit maps of the 120~GeV proton beam for two consecutive 12-hour overnight data-taking runs due to limited beam time at the FTBF test facility.
The maps represent the number of matched hits in the column–row coordinates of the AstroPix sensor.
The sensor surface was aligned perpendicular to the beam axis, so the column and row directions directly correspond to the transverse $x$- and $y$-directions of the incident beam.
To quantify the beam profile, each two-dimensional map was projected onto the column and row axes.
By multiplying the column and row indices by the pixel pitch of 500~\unit{\micro\meter}, the projections were converted into physical $x$- and $y$-coordinate histograms of the number of matched hits.
Gaussian functions were fitted to these one-dimensional distributions to extract the beam widths along each axis.

Figure~\ref{fig:beamprofile} shows a representative example of the fitting result for the extracted beam profile.
From Gaussian fits in $x$ and $y$, the beam profiles for each of the three runs were measured to be
$\sigma_{x} \times \sigma_{y} =$ 4.3~mm $\times$ 3.5~mm and 4.3~mm $\times$ 3.9~mm, respectively.
As mentioned in Section~\ref{sec:ftbf}, in the absence of MWPC-based tracking, the extracted beam profiles could not be directly compared with an independent reference measurement~\cite{Klest_2025}.
However, the measured beam widths are consistent with the expected FTBF beam conditions, which specify a proton beam spot size confined within approximately 6 mm, as described in the previous section.

\begin{figure*}[h!]
    \centering
    \includegraphics[width=0.95\textwidth]{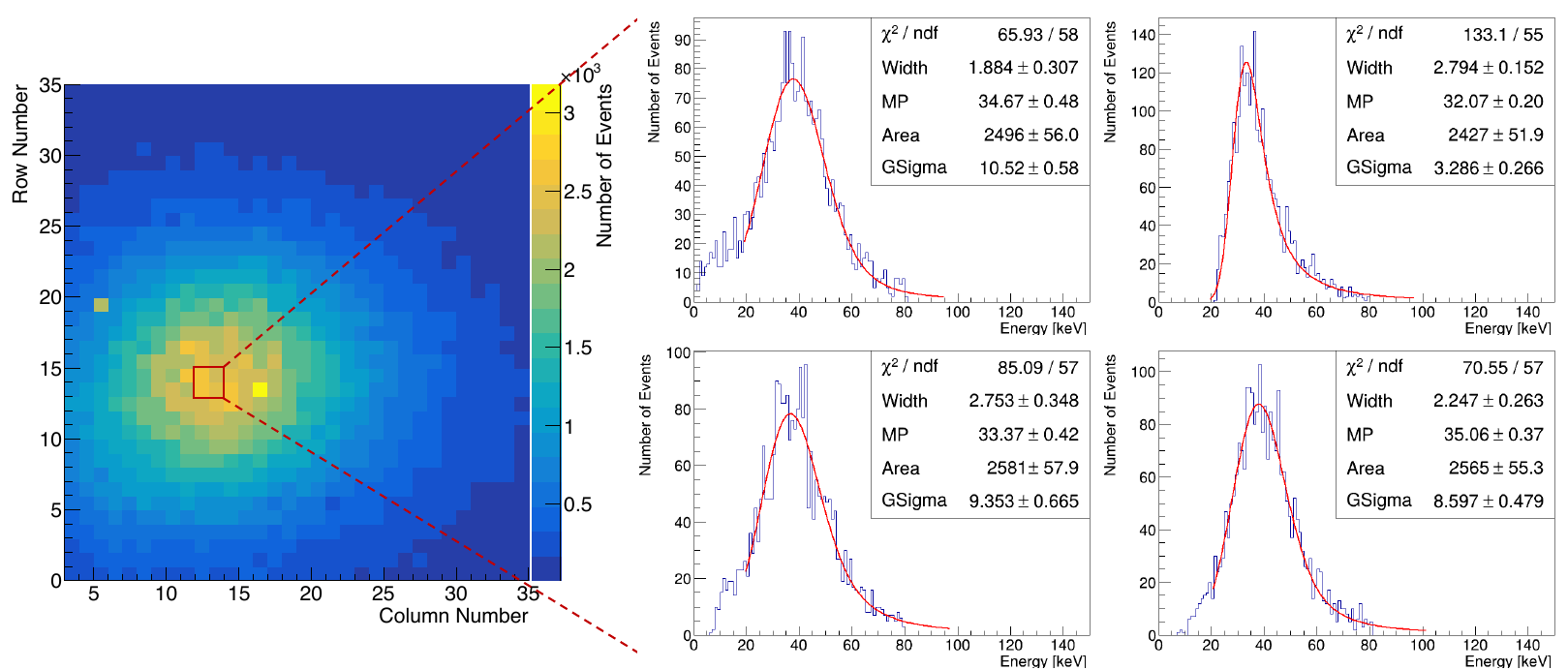}
    \caption{(Left) Example hit map of 120 GeV proton run. (Right) Example fit energy distributions for 2 $\times$ 2 pixels at the beam center after calibration. All spectra have been fitted with a Landau function convoluted with a Gaussian function, and the fit parameters are presented in the legend.}
    \label{fig:protonfit}
\end{figure*}

\subsection{Energy Measurement}

In the 120 GeV proton data analysis, only pixels with more than 1,000 matched hits were used to ensure sufficient statistics for each run. 
These pixels are predominantly located in the beam center region. 
For each selected pixel, the ToT distribution was converted into an energy distribution using a pixel-by-pixel calibration curve.
The resulting energy distribution was fitted with a Landau function convoluted with a Gaussian function, and the Most Probable Value (MPV) was extracted from the fit.
Figure~\ref{fig:protonfit} shows a representative two-dimensional hit map and the corresponding fitted energy distributions for a $2 \times 2$ pixel region at the beam center, along with the associated fit parameter values.

\begin{figure}[h!]
    \centering
    \includegraphics[width=0.99\linewidth]{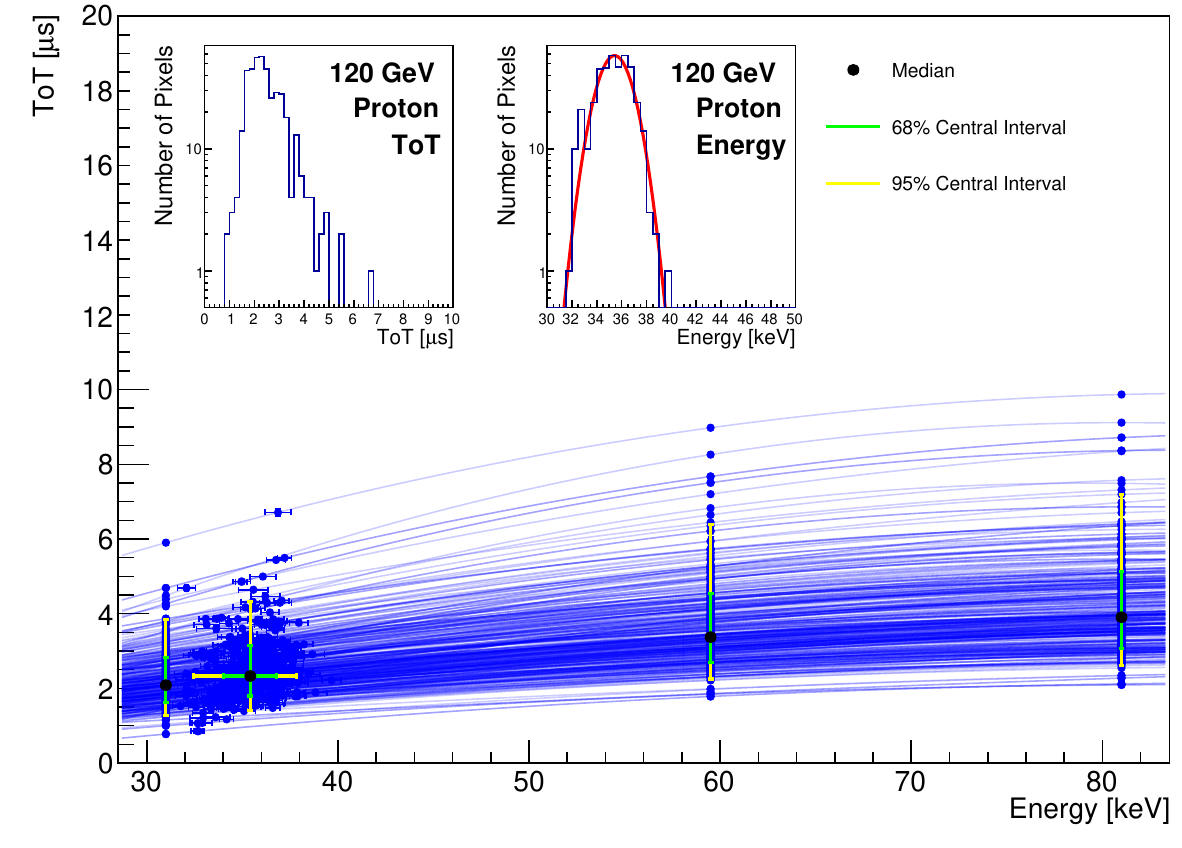}
    \caption{Calibration curves for all selected pixels as a function of energy, including proton measurement results. 
    The upper subplots in each plot show the ToT and energy distributions of the measured 120~GeV protons, respectively.}
    \label{fig:calibwproton}
\end{figure}

Figure~\ref{fig:calibwproton} shows the resulting MPVs from these pixels at the beam center for two runs, plotted on the calibration curve. 
All MPV points fall within the energy range between 31~keV and 59.5~keV, as expected.
The central intervals of the measured MIP energy, corresponding to 68\% (green) and 95\% (yellow) levels, are also overlaid.
The upper panels of the Figure~\ref{fig:calibwproton} show the ToT and energy distributions of the measured 120~GeV protons.
The ToT distribution shape for 120 GeV protons exhibits a pixel-to-pixel variation comparable to those observed for 31, 59.5, and 81~keV calibration points, as shown in the left subpanel of Figure~\ref{fig:calibcurve}, while the corresponding energy distribution follows a Gaussian shape, as illustrated in the right subpanel of Figure~\ref{fig:calibwproton}.

\begin{figure}[h!]
    \centering
    \includegraphics[width=0.8\linewidth]{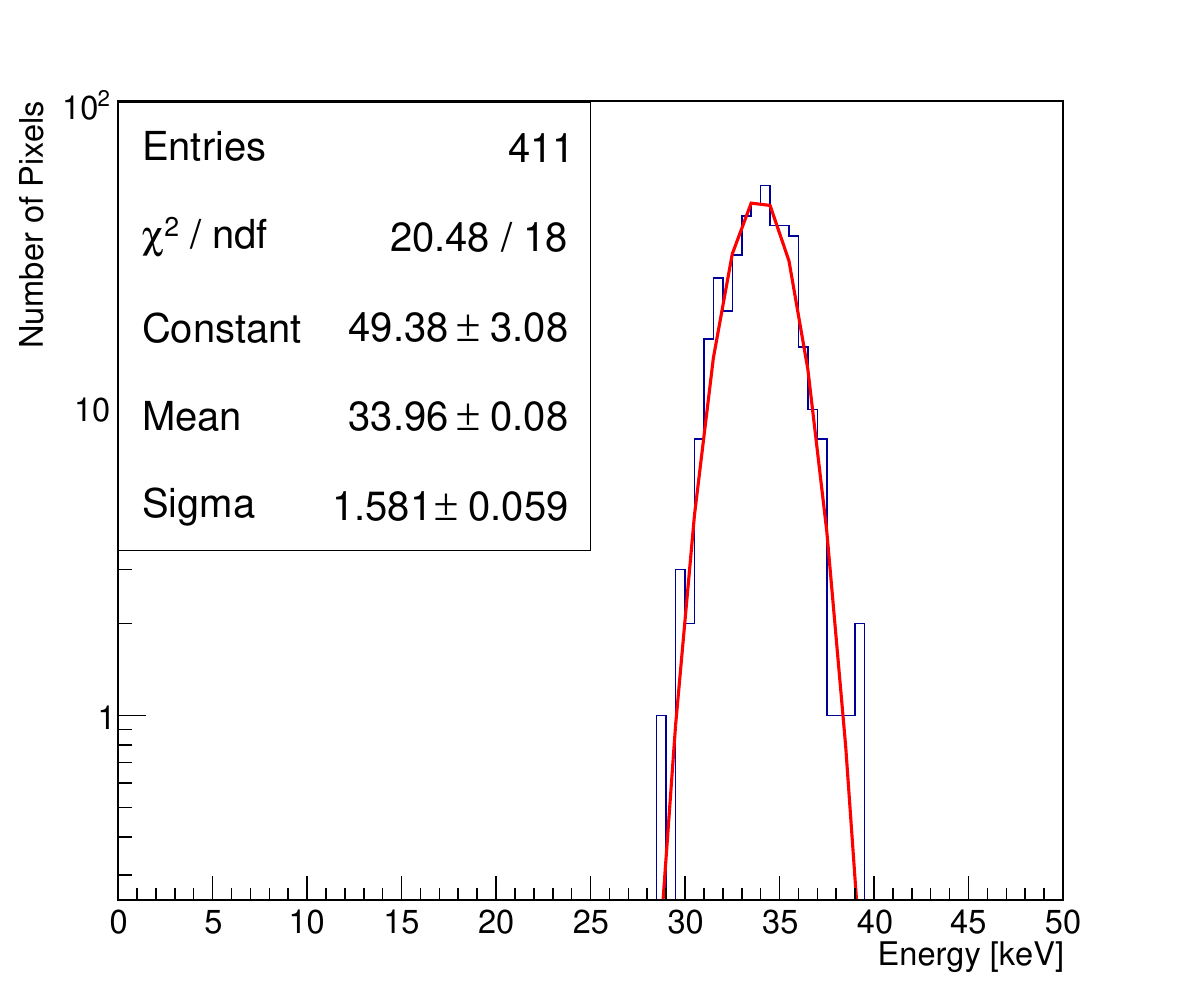}
    \caption{Distribution of the pixel-by-pixel MPV energy obtained from the measured 120~GeV proton data. Each entry corresponds to the MPV extracted from an individual pixel. The red line and the statistics box indicate the Gaussian fit to the distribution.}
    \label{fig:protonenergy}
\end{figure}

Figure~\ref{fig:protonenergy} shows the distribution of the pixel-by-pixel MPV energies obtained from the measured 120~GeV proton data, together with the corresponding Gaussian fit result.
The figure presents calibration results for a total of 411 pixels, all of which are located in rows 21 and below. 
Each entry corresponds to the MPV extracted from an individual pixel.
The MPV distribution has a mean value of 34.0~keV with a sigma of 1.58~keV, corresponding to a relative spread of 4.7\%.
This spread reflects the combined effect of statistical fluctuations and pixel-to-pixel variations, and is therefore taken as the statistical uncertainty of the measurement.
The run-by-run fluctuations of about 2.8\% is consistent with this value.
The pixel-to-pixel variation predominantly reflects differences in the effective depletion depth, which may be attributed to process-induced defects and local pixel-level bias voltage variations.

To evaluate systematic uncertainties, four sources are considered.

The first source arises from the uncertainty in the MPV extracted from a Landau function convoluted with a Gaussian function, fitted independently for each pixel. 
The pixel-by-pixel relative uncertainties show a mean value of 1.5\% with a spread of 0.4\%, with an upper range of 3\%, reflecting pixel-to-pixel variations.
The MPV fit uncertainty, propagated to the final calibrated energy value, is taken to be 3\%. 

As a second source, the calibration uncertainty is considered by propagating the covariance matrix of the calibration parameters for each pixel, resulting in a systematic uncertainty of 1.3\%.
Charge loss due to diffusion into neighbouring pixels is not considered as a source of systematic uncertainty, as charge sharing is expected to be negligible owing to the large pixel size.
Temperature effects are not expected to introduce a relevant systematic uncertainty, as the beam test was performed under stable room-temperature conditions with continuous temperature monitoring and no significant fluctuations observed during data taking.

The momentum spread of the 120 GeV proton beam is negligible at the level of $\Delta p/p \sim$ 2\%, as reported in the FTBF documentation~\cite{ftbfBeamMTest}. 
This corresponds to a variation in $\beta\gamma$ from approximately 125 to 130, which results in no observable change in the most probable energy loss in silicon over this range and is therefore not considered an additional systematic effect.

The average MPV of the MIP energy, obtained from pixel-by-pixel measurements, including systematic uncertainties, is therefore 34.0 $\pm$ 1.9~keV.
It corresponds to an effective depletion depth of 128 $\pm$ 8~\unit{\micro\meter} as derived from the most probable energy loss calculation for 120~GeV protons in silicon~\cite{ParticleDataGroup, RevModPhys.60.663}.

The extracted depletion depth should be interpreted as an effective depletion depth for MIP signals under the AstroPix\_v3 operating conditions. 

The calibrated MIP peak at 34~keV lies above the lowest calibration point at 31~keV. 
If interpreted as a MIP energy loss, 31~keV would correspond to an effective depletion depth of approximately 118~\unit{\micro\meter}. 
This value therefore provides a lower bound for the effective depletion depth inferred from the MIP signal, supporting the extracted value of 128 $\pm$ 8~\unit{\micro\meter} as a physically reasonable estimate.

\section{Conclusions \& Outlook}
AstroPix\_v3 was tested with the 120~GeV proton beam at the Fermilab Test Beam Facility for R\&D purposes to study its MIP response and to measure the energy deposited by MIP and the corresponding effective depletion depth.
The energy distributions from the 120 GeV protons were well described by a Landau function convoluted with a Gaussian function, from which the MPV of deposited MIP energy was extracted on a pixel-by-pixel basis.
The average MPV of MIPs across all pixels at the beam center and over all runs was measured to be 34.0 $\pm$ 1.9~keV, including systematic uncertainties, corresponding to the effective depletion depth of 128 $\pm$ 8~\unit{\micro\meter}.
The MIP energy deposition lies well within the AstroPix\_v3 dynamic range of 25--200~keV, demonstrating the suitability of the sensor for MIP detection.
A relative pixel-to-pixel variation of 4.7\% was observed, reflecting a corresponding variation in the effective depletion depth across the sensor and indicating good uniformity. 

This work provides the first experimental validation that the AstroPix\_v3 can detect and characterize MIP signals under realistic conditions.
It is intended for use in space-based experiments under similar operating conditions, including the same bias voltage, and has been selected as the baseline sensor for the imaging layer of the BIC.
Therefore, the measured depletion depth at $-150~\mathrm{V}$ represents a physically relevant operating point for the detector.
The extracted depletion depth serves as an experimentally validated input for detector simulations, enabling more realistic modeling of energy deposition in both space-based and ePIC experiments.
Given its intended use, independent performance validation is particularly important, and the demonstrated MIP response provides a reliable basis for calibration.

Although AstroPix\_v3 is not the final design for the BIC in the ePIC experiment at the EIC, it serves as a valuable platform for performance validation and system-integration studies.
Current R\&D has advanced to multi-layer, daisy-chained multi-chip operation tests using quad-chips and nine-chip prototype modules, successfully demonstrating the proof of principle for the fundamental BIC imaging-layer unit~\cite{kim2025performanceastropixprototypemodule}.

The later versions mainly focus on incremental performance improvements and application-specific optimization.
AstroPix\_v4~\cite{suda2026astropix4, striebig2024astropix4} introduces individual pixel readout and pixel-by-pixel threshold tuning DAC in a smaller reticle format, while maintaining the same fundamental chip design parameters as AstroPix\_v3.
AstroPix\_v5 is a full-scale version of AstroPix\_v4 for the ComPair2 balloon flight, including bug fixes and a fabrication-foundry switch, whereas AstroPix\_v6 is an application-specific size variant developed for the BIC.
Importantly, the fundamental sensor and chip design concepts remain largely unchanged across these versions. Therefore, the performance studies of AstroPix\_v3 remain directly relevant and provide essential feedback for the development and optimization of future versions.

\section*{Acknowledgments} 
Development of AstroPix is funded through the following NASA programs: 18-APRA18-0084 and 20-RTF20-0003.
This document was prepared by the members of the Barrel Imaging Calorimeter (BIC) Detector Subsystem Collaboration (DSC) of the ePIC Collaboration using the resources of the Fermi National Accelerator Laboratory (Fermilab), a U.S. Department of Energy, Office of Science, Office of High Energy Physics HEP User Facility. 
Fermilab is managed by FermiForward Discovery Group, LLC, acting under Contract No. 89243024CSC000002.
The material is based upon work partially supported by the U.S. Department of Energy, Office of Science, Office of Nuclear Physics and Laboratory Directed Research and Development (LDRD) funding from Argonne National Laboratory, provided by the Director, Office of Science, of the U.S. Department of Energy under Contract No. DE-AC02-06CH11357; the Electron-Ion Collider Project R\&D Funds for the eRD115 Project.
We also acknowledge the invaluable support of the FTBF staff, whose expertise and dedication greatly facilitated the execution of our test beam experiments.

\bibliographystyle{elsarticle-num}
\bibliography{bib_rev.bib}

\end{document}